\documentclass[sigconf]{aamas}

\usepackage{balance} 
\usepackage{multirow}

\doi{OITH7375}

\makeatletter
\gdef\@copyrightpermission{
  \begin{minipage}{0.2\columnwidth}
   \href{https://creativecommons.org/licenses/by/4.0/}{\includegraphics[width=0.90\textwidth]{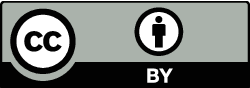}}
  \end{minipage}\hfill
  \begin{minipage}{0.8\columnwidth}
   \href{https://creativecommons.org/licenses/by/4.0/}{This work is licensed under a Creative Commons Attribution International 4.0 License.}
  \end{minipage}
  \vspace{5pt}
}
\makeatother

\setcopyright{ifaamas}
\acmConference[AAMAS '26]{Proc.\@ of the 25th International Conference
on Autonomous Agents and Multiagent Systems (AAMAS 2026)}{May 25 -- 29, 2026}
{Paphos, Cyprus}{C.~Amato, L.~Dennis, V.~Mascardi, J.~Thangarajah (eds.)}
\copyrightyear{2026}
\acmYear{2026}
\acmDOI{}
\acmPrice{}
\acmISBN{}

\title[AAMAS-2026 Formatting Instructions]{Beyond Self-Interest: Modeling Social-Oriented Motivation \\ for Human-like Multi-Agent Interactions}

\author{Jingzhe Lin}
\affiliation{
  \institution{
    School of Artificial Intelligence,  \\ 
        Beijing Normal University \\
    \& Beijing Key Laboratory of Artificial Intelligence for Education\\
    \& Engineering Research Center of Intelligent Technology and Educational Application, MOE\\   
  }
  \city{Beijing}
  \country{China}
}
\email{linjingzhe@bnu.edu.cn}

\author{Ceyao Zhang}
\affiliation{
  \institution{
    School of Intelligence Science\\ and Technology,\\ Institute for Artificial Intelligence,\\ Peking University
  }
  \city{Beijing}
  \country{China}
}
\email{ceyaozhang@pku.edu.cn}

\author{Yaodong Yang}
\affiliation{
  \institution{
    Institute for Artificial Intelligence, Peking University
  }
  \city{Beijing}
  \country{China}
}
\email{yaodong.yang@pku.edu.cn}

\author{Yizhou Wang}
\affiliation{
  \institution{
  School of Computer Science,\\ Institute for Artificial Intelligence, \\  State Key Laboratory of General Artificial Intelligence, \\ Peking University\\
  }
  \city{Beijing}
  \country{China}
}
\email{yizhou.wang@pku.edu.cn}

\author{Song-Chun Zhu}
\affiliation{
  \institution{
    School of Intelligence Science and Technology, Institute for Artificial Intelligence,\\ Peking University \\
    \& 
    Beijing Institute for General Artificial Intelligence
  }
  \city{Beijing}
  \country{China}
}
\email{s.c.zhu@pku.edu.cn}

\author{Fangwei Zhong}
\authornote{Corresponding author.}
\affiliation{
  \institution{
    School of Artificial Intelligence,  \\ 
        Beijing Normal University \\
    \& Beijing Key Laboratory of Artificial Intelligence for Education\\
    \& Engineering Research Center of Intelligent Technology and Educational Application, MOE\\   
  }
\city{Beijing}
  \country{China}
}
\email{fangweizhong@bnu.edu.cn}

\begin{abstract}
Large Language Models (LLMs) demonstrate significant potential for generating complex behaviors, yet most approaches lack mechanisms for modeling social motivation in human-like multi-agent interaction.
We introduce Autonomous Social Value-Oriented agents (ASVO), where LLM-based agents integrate desire-driven autonomy with Social Value Orientation (SVO) theory. At each step, agents first update their beliefs by perceiving environmental changes and others' actions.
These observations inform the value update process, where each agent updates multi-dimensional desire values through reflective reasoning and infers others' motivational states. By contrasting self-satisfaction derived from fulfilled desires against estimated others' satisfaction, agents dynamically compute their SVO along a spectrum from altruistic to competitive, which in turn guides activity selection to balance desire fulfillment with social alignment.
Experiments across School, Workplace, and Family contexts demonstrate substantial improvements over baselines in behavioral naturalness and human-likeness. These findings show that structured desire systems and adaptive SVO drift enable realistic multi-agent social simulations.
\noindent\textbf{Project page: } 
\url{https://asvo-agents.github.io/ASVO-agents/}
\end{abstract}


\newcommand{\BibTeX}{\rm B\kern-.05em{\sc i\kern-.025em b}\kern-.08em\TeX}

\begin{document}


\pagestyle{fancy}
\fancyhead{}


\maketitle 


\section{Introduction}

Simulating human-like behavior has long been a core goal in artificial intelligence~\cite{fetzer1990artificial}. Beyond reproducing individual actions, it is important to model the complex social dynamics in realistic multi-agent scenarios~\cite{becker1974theory,shu2020adventures}, where agents are endowed with diverse personalities, motivations, and social preferences. Such simulations have potential applications in domains including education~\cite{zhang-etal-2025-simulating}, transportation~\cite{doi:10.1073/pnas.1820676116}, and finance~\cite{yang2025twinmarketscalablebehavioralsocial}. 

Recent large language models~\cite{achiam2023GPT,cheng2024exploring} have opened a new era of agent-based social simulation. Pioneering frameworks such as Generative Agents~\cite{10.1145/3586183.3606763} and ProAgent~\cite{zhang2024proagent} demonstrate autonomous social behaviors, while subsequent work extends this paradigm to large-scale reasoning~\cite{li2023camel}, institutional settings~\cite{li2024simhospital}, or general social simulation~\cite{zhang2024yulan, wang2024concordia}. These advances reveal that LLMs enable complex social inference and emergent norms~\cite{Binz_2023}, yet most current approaches rely on static~\cite{wang2024a} or rule-based agent~\cite{yao2023react} personalities, limiting their ability to capture the dynamics of human intrinsic motivation.

Consequently, there is growing interest in integrating structured psychological constructs such as \textbf{Social Value Orientation (SVO)}~\cite{murphy2011measuring} and motivational models into LLM agents~\cite{zhang2023heterogeneous,knight2023alignment,kirk2023value,wang2025desire}. Recent progress in \textbf{desire-driven autonomy} shows agents better emulate human-like behavior when guided by intrinsic motivations rather than fixed goals~\cite{wang2025desire}. However, these approaches primarily focus on task generation and lack mechanisms to capture how social preferences shape complex multi-agent interactions. This is a gap this paper addresses by coupling desire systems with adaptive SVO computation.

\begin{figure}[t]
\centering
\vspace{-6pt}
\includegraphics[width=0.98\linewidth,trim=0 6 0 5,clip]{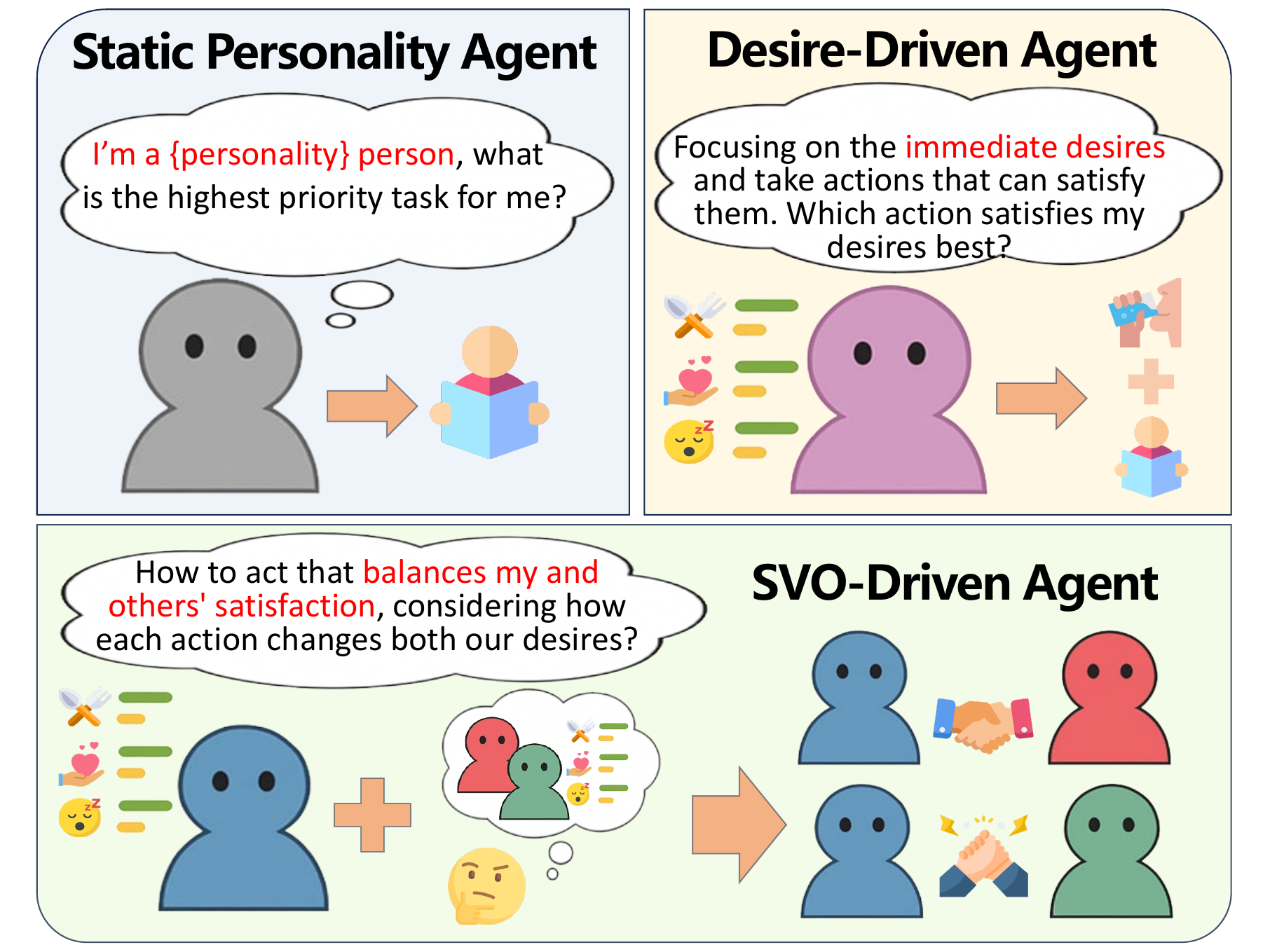}
\vspace{-5pt}
\caption{Upper-left: Static Personality agents condition actions on fixed goals.
Upper-right: Desire-Driven agents produce human-like behavior by grounding decisions in internal desires.
Lower: SVO-Driven agents (ours) build upon desire-based autonomy by incorporating Social Value Orientation and explicitly weighing each action’s impact on both self and others’ anticipated satisfaction, thereby moving beyond pure self-interest toward a richer model of human social behavior.}
\label{fig:teasor}
\vspace{-9pt}
\end{figure}

In this paper, we propose \textbf{Autonomous Social Value-Oriented agents (ASVO)}, a novel framework that integrates desire-driven autonomy with social motivation grounded in SVO. Within this framework, agents are categorized into four core social personality types: \textit{Altruistic}, \textit{Prosocial}, \textit{Individualistic}, and \textit{Competitive}. Each type reflects a distinct trade-off between self-interest and consideration for others. This unified framework also enables agents to adaptively infer their own and others’ motivations, engage in complex social interactions, and demonstrate value shifts that mirror complex human dynamics. Built on the Concordia~\cite{wang2024concordia}, our method enables agents to actively take activities to update and satisfy both intrinsic desires and social value preferences as they interact with others in diverse simulated societies. 

The contributions are threefold:
1) We introduce an LLM-based framework, ASVO, which embeds Social Value Orientation into desire-driven agents, facilitating dynamic motivational adaptation in multi-agent environments.
2) We design a multi-dimensional evaluation metric to strictly assess behavioral alignment, covering cooperation dynamics, naturalness, and human-likeness.
3) We validate ASVO through extensive benchmarking across three distinct contexts, showing that it yields more interpretable and socially consistent behaviors compared to strong baselines.

\begin{figure*}[t]
\centering
\vspace{-7pt}
\includegraphics[width=0.92\textwidth,trim=0 35 0 3,clip]{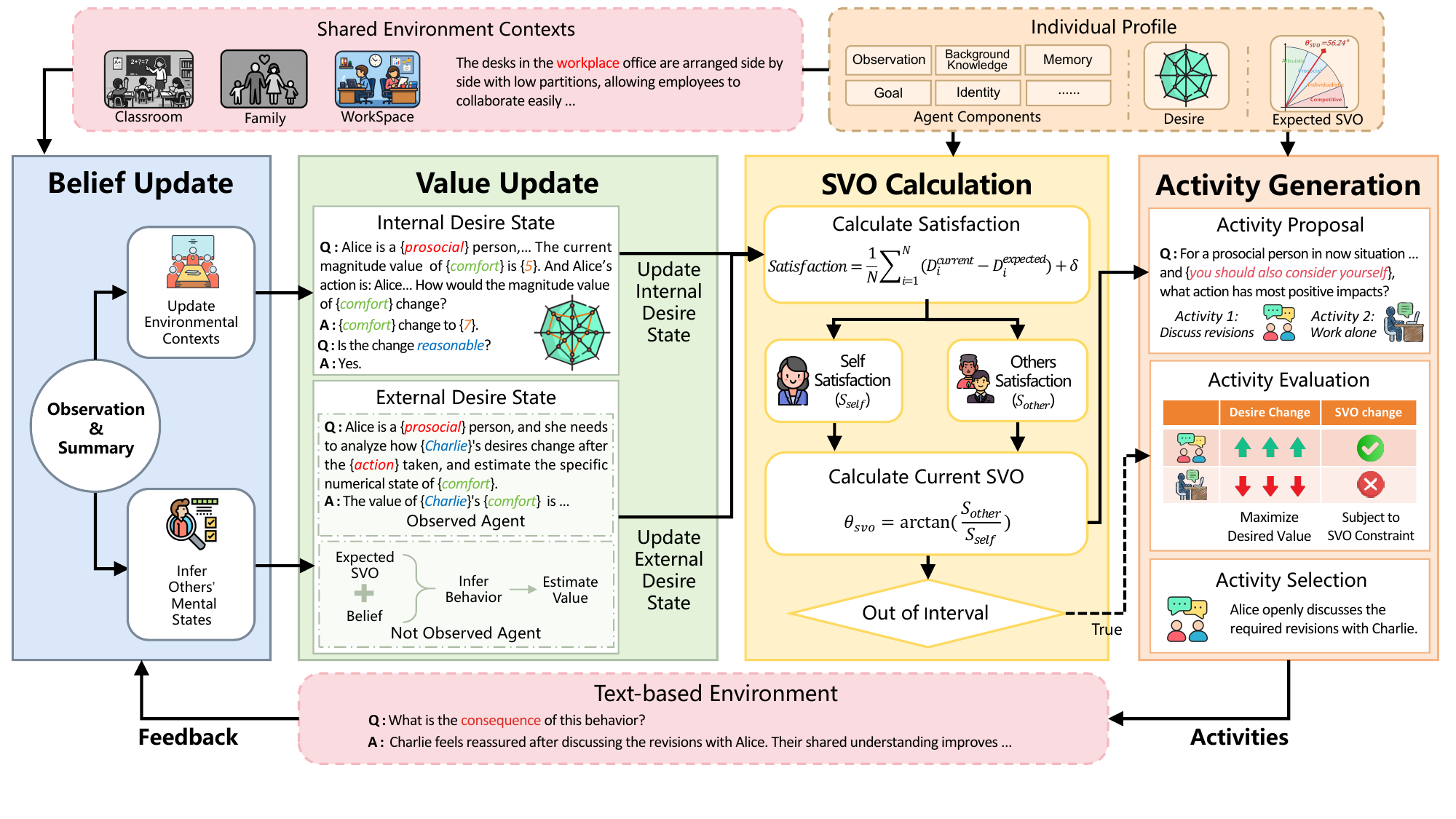}
\vspace{-5pt}
\caption{Overview of the ASVO framework. During the \textbf{Initialization} stage, each agent’s \textbf{Individual Profile} is constructed, specifying its social personality type (expected SVO), a set of core psychological desires, and agent-specific components including \emph{Observation}, \emph{Memory}, \emph{Background Knowledge}, \emph{Identity}, and \emph{Goal}. These attributes are encoded into the agent’s internal state, forming the basis of simulation. At each simulation step, the pipeline iteratively proceeds through four key modules: \textbf{Belief Update} (agents incorporate shared environment contexts and summarize observations to update beliefs about others’ states), \textbf{Value Update} (updating the Internal Desire State via self-reflection and the External Desire State by estimating others’ needs, including unobserved agents), \textbf{SVO Calculation} (computing self and others’ satisfaction to compute the agent’s current SVO with an interval check), and \textbf{Activity Generation} (proposing, evaluating, and selecting LLM-driven activities by synthesizing desires, context, and SVO). This forms a closed feedback loop over a text-based environment, enabling the co-evolution of motivations and social values throughout multi-agent social interaction.}
\vspace{-6pt}
\label{fig:framework}
\end{figure*}

\section{Related Work}

\paragraph{\textbf{Social Simulation with Large Language Models.}} Recent advances in large language models (LLMs) have improved the realism and complexity of agent-based social simulations~\cite{ziheng2025llm, xu2025comet}. Pioneering frameworks such as Voyager~\cite{xu2023voyager} and CAMEL~\cite{li2023camel} showed that LLM-driven agents can autonomously explore, communicate, and collaborate in virtual environments, enabling complex group behaviors, norms, and roles. Further, earlier intrinsic-motivation-driven LLM agents, such as D2A~\cite{wang2025desire}, together with simulators like Ecoagent~\cite{ecoagent2024}, OASIS~\cite{oasis2023}, and AgentScope~\cite{agentscope2024}, have extended simulation capabilities toward larger-scale and more diverse social environments. Subsequent research has expanded into domains including social dilemma games~\cite{leibo2017multi}, online community simulation~\cite{andrei2023societal}, and value alignment in digital societies~\cite{gabriel2023artificial}. These works demonstrate that LLMs enable complex language interaction and adaptive behavior modeling, pushing the boundaries of computational social science. However, most existing LLM-based simulations rely on static or rule-based agent personalities, which fail to capture the full spectrum of human-like heterogeneity and dynamic value change. Thus, there is growing interest in integrating structured psychological constructs, such as Social Value Orientation (SVO) and motivational models, into LLM agents to simulate real-world social phenomena~\cite{zhang2023heterogeneous,knight2023alignment,kirk2023value}.

\vspace{-4pt}
\paragraph{\textbf{Social Value Orientation in Agent-Based Models.}} Social Value Orientation (SVO) is a key concept for quantifying individual differences in social preferences, cooperation, and competition~\cite{murphy2011measuring,vanlange2021social}. SVO classifies agents along a continuum from altruistic and prosocial to individualistic and competitive types, providing a grounded metric for modeling how agents weigh their own welfare against that of others. Continuous SVO measures and related constructs have become standard tools for psychological assessment and multi-agent simulations, enabling fine-grained analysis of resource allocation, group coordination, and social dilemmas~\cite{grzelak2022svo}. Recent work on multi-agent systems has shown that assigning heterogeneous SVO values to agents leads to more realistic social outcomes. For example, in human-robot teams, SVO-inspired reward shaping improves fairness, trust, and group performance~\cite{williams2021modeling}. In computational sociology, SVO-based models enable the study of emergent cooperation, norm formation, and polarization, especially when combined with networked agent learning~\cite{macy2002social}. Despite these advances, most prior work treats SVO as a static attribute, failing to capture the dynamic and context-sensitive nature of human social preferences. Recent work on desire-driven autonomy~\cite{wang2025desire} enables agents to evolve intrinsic motivations in response to experience and feedback, offering new opportunities for modeling personality plasticity and adaptive decision-making in multi-agent systems.

\vspace{-4pt}
\paragraph{\textbf{Motivation and Personality in LLM Agents.}}
Modeling dynamic motivation and personality remains a central challenge for AI-based social simulation. Traditional multi-agent systems often rely on handcrafted rules or fixed reward structures, limiting their expressiveness and adaptability~\cite{epstein2006generative}. The integration of LLMs allows richer language understanding and social inference, but explicit mechanisms for intrinsic motivation, needs satisfaction, and personality evolution are still lacking. Recent approaches bridge this gap by introducing dual motivational frameworks, combining classic psychological theories such as Maslow’s hierarchy of needs~\cite{maslow1943theory}, Self-Determination Theory~\cite{ryan2000self}, or Big Five personality modeling~\cite{john1999bigfive} with SVO and value reasoning. These frameworks support more realistic simulation of individual adaptation, group influence, and value-driven decision-making in open-ended social environments~\cite{wang2025desire}.

\section{Preliminaries}

\paragraph{\textbf{Social Value Orientation (SVO)}~\cite{murphy2011measuring}} is a psychological construct that quantifies how individuals balance their own interests with those of others during social interactions. 
It captures individual differences in social preferences across people and contexts, reflecting how much weight one places on others' welfare when making decisions. 
Accordingly, it serves as a foundational framework for explaining cooperative, altruistic, and competitive behaviors in groups. 
Formally, SVO is represented as an angle:
\begin{equation}
    \theta_{SVO} = \arctan\left( \frac{V_{\mathrm{other}}}{V_{\mathrm{self}}} \right),
\end{equation}
where $V_{\mathrm{self}}$ and $V_{\mathrm{other}}$ denote the values attributed to oneself and to others, respectively. Based on this angle, individuals are categorized into four types: \textit{altruistic} (high concern for others), \textit{prosocial} (balancing self and others), \textit{individualistic} (focused on self-interest), and \textit{competitive} (maximizing relative advantage). This classification provides a theoretical foundation for modeling diverse human behaviors and has been widely applied in multi-agent systems~\cite{collins2023social}, behavioral economics~\cite{van1999pursuit}, and social psychology~\cite{van1997development}, enabling realistic simulation of social norms and collective decision-making.

\section{Problem Formulation}

In this work, we formalize the problem of modeling socially adaptive agents that integrate structured intrinsic motivations with dynamically evolving social value orientations (SVOs). 
Each agent continuously balances self- and other-oriented drives in a shared environment, allowing value-based reasoning and context-sensitive social behavior to emerge. 

Let $s_t$ denote the environment state at time $t$, and $o_{0:t-1}$ and $a_{0:t-1}$ represent the sequences of past observations and actions, respectively. 
Each agent $i$ is characterized by an internal desire vector $D_t^i$, a social value orientation angle $\theta_t^i$, and a fixed attribute set $\psi_i$ that encodes static features such as identity, memory, and personality. 
At each step, the agent generates a new activity conditioned on its experience, motivational state, and social orientation:
\begin{equation}
a_t^i \sim \textsc{Agent}(\,\cdot \mid o_{0:t-1},\, a_{0:t-1},\, D_t^i,\, \theta_t^i;\, \psi_i\,),
\label{eq:agent}
\end{equation}
Here, $D_t^i$ encodes the agent’s intrinsic motivations across multiple dimensions, while $\theta_t^i$ determines how self- and other-oriented satisfactions are weighted when evaluating possible actions. 
The static attributes $\psi_i$ serve as contextual priors that shape consistent identity and reasoning style throughout the simulation. 
Together, these components define a unified decision-making process in which intrinsic desires interact with social preferences to produce coherent behaviors. 
The agent selects actions that maximize its desire satisfaction while ensuring the resulting SVO remains within its expected personality range, i.e., $a_t^i = \arg\max_{a} U_{\text{self}}(a \mid D_t^i) \text{ s.t. } \theta_t^i(a) \in [\theta_{\min}^i, \theta_{\max}^i]$.

\section{Methodology}
In this section, we will explain each module of our multi-agent simulation in detail, as shown in Figure~\ref{fig:framework}.
We first initialize the individual profile and environmental contexts. 
The core simulation consists of four modules: (1) \textbf{belief updates}, (2) \textbf{value updates}, (3) \textbf{SVO calculations}, and (4) \textbf{activity generation}.
Lastly, we describe how each module iterates.

\subsection{Individual Profile Initialization}
At the start of each simulation, a set of agents is instantiated within a shared environment context. Each agent is initialized with an \textit{expected SVO}, representing its initial social personality type. During simulation, this expected SVO serves as a reference, while the agent's \textit{current SVO} is dynamically updated at each step based on ongoing behavioral feedback and social context. Thus, the agent's social value orientation is not static, but adaptively evolves throughout the simulation process.

In addition to SVO, each agent is equipped with an individual profile that includes eight structured desire dimensions forming the foundation of its intrinsic motivation system. These desires, together with the agent’s background knowledge, memory, and identity information, constitute the initial cognitive and affective states used by the reasoning modules. 

This dual-SVO and multi-desire initialization provides a psychologically grounded motivational baseline that shapes each agent’s value tendency and perception of others. It also prepares subsequent modules such as belief update and activity generation, ensuring that social adaptation arises from coherent internal dynamics.

\subsection{Belief Update}
\label{subsec:belief-update}
The \textbf{belief update} module serves as a core process in our framework, underpinning both self-modeling and modeling of others. 
At each step, agents actively perceive changes in the environment independently and update their internal belief of the state, denoted as $b_t^i$.Specifically, agents gather new observations about:
1) \textbf{Environmental Contexts}: Changes in shared environment states, such as shifts in group membership or resources, represented as $o_t^{env}$;
2) \textbf{Others' Actions}: Observations or inferences about other agents’ actions, represented as $o_t^{other}$.  
These beliefs together provide the foundation for subsequent modeling of both self and others.

\subsection{Value Update}
At each simulation step, each agent updates both its internal desire state and its estimation of others’ desire states. 

\textbf{Internal Desire State:} 
Each agent maintains a structured, multi-dimensional desire state to capture evolving needs during interaction. 
It is updated via an LLM-mediated reflective loop: the agent’s most recent action and observed consequences are fed into a structured prompt, with the agent’s SVO imposed as an explicit constraint. 
For example, a prosocial agent like Alice organizing a review session weighs personal comfort against collective benefit, aligning choices with both individual and group welfare. 
An update is accepted only if it remains consistent with the agent’s established SVO profile; otherwise, the model iterates reflection and correction. 
Specifically, in this paper we instantiate the desire state with eight dimensions, while the mechanism generalizes beyond this choice.

\textbf{External Desire State:}
To estimate others’ desires, agents perform SVO-conditioned social inference. 
If the target agent is visible (e.g., Charlie), the agent directly infers their desire state from behavioral cues; if invisible (e.g., Bob), it first infers the most likely behavior from SVO and contextual memory, then estimates the current desire state conditioned on that behavior. 
This aligns with the Value Update pathways in Figure~\ref{fig:framework}. 
Formally, given $o_{0:t-1}$ and $a_{0:t-1}$, agent $i$ infers the latent motivational state of agent $j$ via:
\begin{equation}
\widehat{D}_t^{\,j|i} \sim \text{LLM}(o_{0:t-1},\, a_{0:t-1},\, \theta_t^i),
\label{eq:belief-update}
\end{equation}
where $\widehat{D}_t^{\,j|i}$ denotes $i$’s estimate of $j$’s desire vector, and $\widehat{\theta}_t^{\,j|i}$ denotes the inferred SVO of $j$. 
This step turns observations into socially meaningful representations, enabling anticipation of others’ intentions for subsequent value updates and decision-making.

\subsection{SVO Calculation}
Throughout the simulation, each agent maintains both a fixed \textit{expected SVO} (personality baseline) and a \textit{dynamic current SVO} that is continuously recalculated based on recent social interactions and value satisfaction. The current SVO serves as the immediate basis for decision-making and is regularly compared to the expected SVO to regulate behavioral consistency and personality alignment.

Following value updates, each agent calculates its current SVO by comparing internal and external satisfaction. For $N$ tracked desires, satisfaction is computed as the average difference between the current and expected values:
\begin{equation}
S \;=\; \frac{1}{N}\sum_{i=1}^{N}\big(D^{\text{current}}_i - D^{\text{expected}}_i\big) + \delta,
\end{equation}
where $D^{\text{current}}_i$ and $D^{\text{expected}}_i$ represent the present and expected values of the $i$-th desire, respectively (in our implementation, each desire takes values in $[0,10]$). In practice, we clip the per-dimension differences $\big(D^{\text{current}}_i - D^{\text{expected}}_i\big)$ to $[-4,4]$, with this numeric range chosen based on empirical experience to stabilize the ratio. This bounded design also reflects the fact that extreme discrepancies tend to saturate and no longer change perceived satisfaction~\cite{kahneman1979prospect}. Here $\delta$ is a constant offset chosen to shift satisfactions to be positive before taking the ratio, ensuring a stable and well-defined SVO angle.
The SVO angle is then determined as:
\begin{equation}
\theta_{svo} = \arctan\Big( \frac{S_{\mathrm{other}}}{S_{\mathrm{self}}} \Big),
\end{equation}
where $S_{\mathrm{self}}$ and $S_{\mathrm{other}}$ are the satisfaction scores for self and others.

If the computed SVO value deviates from the expected personality interval (for example, if a prosocial agent's SVO is unusually high or low), a regulatory mechanism is triggered at the activity generation stage. The agent’s action selection prompt is then adaptively modified to guide its choices gently back toward the expected SVO range. This iterative feedback, illustrated in Figure~\ref{fig:framework}, enables dynamic and interpretable regulation of social value orientation through natural language guidance.

\subsection{Activity Generation}
Based on the regulated SVO, ASVO generates agent behavior through a structured multi-step process that synthesizes motivational needs, social context, and value orientation. At each step, the agent assembles relevant information, including updated desires, contextual memory, and recent observations, into the LLM input, which is adaptively modified by SVO regulation to guide behavior as needed.

The LLM proposes multiple candidate actions based on the agent’s needs and current SVO. Each candidate is evaluated for its predicted impact on self-satisfaction and alignment with value orientation. For example, a prosocial agent may weigh sharing notes (collective benefit) against collecting problems (self-benefit) to determine which better fits its SVO.

Formally, the action selection process can be modeled as a constrained optimization that maximizes the agent’s own desire satisfaction while maintaining consistency with its social value orientation (SVO). 
Let $\mathcal{C}_t^i=\{\tilde a_{t,k}^i\}_{k=1}^K$ denote the candidate action set generated by the LLM at step $t$, and let $U_{\text{self}}(a\mid D_t^i)$ represent the predicted improvement in the agent’s internal desires when taking action $a$. 
Using the SVO Calculation module, the agent predicts the resulting SVO angle for each candidate, denoted $\tilde\theta_t^i(a)$. 
Given the expected SVO interval $[\theta_{\min}^i,\theta_{\max}^i]$, the final action is selected as:
\vspace{-2pt}
\begin{equation}
a_t^i \;=\; \arg\max_{\,a\in\mathcal{C}_t^i}\; U_{\text{self}}(a \mid D_t^i)
\quad \text{s.t.}\quad 
\tilde\theta_t^i(a)\in\big[\theta_{\min}^i,\;\theta_{\max}^i\big].
\end{equation}
\vspace{-2pt}
This formulation ensures that the chosen behavior not only maximizes the agent’s internal satisfaction but also stays within the acceptable SVO range, thereby maintaining a balance between self-driven motivation and social alignment.

The agent selects the action that best balances desire fulfillment and SVO alignment. This chosen action is executed, and the outcomes are used to update desires and SVO in the next step. In this way, agent behavior remains contextually appropriate and value-aligned throughout the simulation, with regulatory adjustments ensuring personality consistency even as social dynamics evolve. Full prompts are provided in the Appendix.

\subsection{Iterative Cycle}
\label{subsec:iter-cyc}

The ASVO framework operates through a continuous perception, motivation, value, action loop. 
This process integrates all preceding modules into a unified and adaptive behavioral mechanism. 
Formally, the overall cycle at time step $t$ can be expressed as:
\vspace{-2pt}
\begin{equation}
s_t \Rightarrow o_{0:t}
\Rightarrow
(D_t^i,\, \widehat D_t^{\,j|i})
\Rightarrow
\theta_t^i
\Rightarrow
a_t^i
\Rightarrow
s_{t+1},
\label{eq:cycle}
\end{equation}
\vspace{-2pt}
where $s_t$ is the environment state, $o_{0:t}$ represents the accumulated observations, $(D_t^i,\, \widehat D_t^{\,j|i})$ are the updated internal and inferred external desire states, $\theta_t^i$ denotes the current social value orientation, and $a_t^i$ is the generated action.

At each iteration, agents first perceive environmental cues and others’ actions to update their beliefs and observation histories $o_{0:t}$. 
They then revise both self- and other-related motivational states $(D_t^i, \widehat D_t^{\,j|i})$ through reflective and inferential reasoning, which jointly shape the updated SVO $\theta_t^i$. 
Conditioned on this updated motivational–social state, the agent selects an action $a_t^i$ according to its policy, balancing self-oriented needs with socially aligned value considerations. 
The resulting action modifies the environment to $s_{t+1}$, providing new contextual feedback for the next step.  

Through this iterative cycle, ASVO agents continuously adapt their desires, infer others’ intentions, and refine their social value orientations. 
This dynamic feedback mechanism ensures that simulated behaviors remain context-sensitive, interpretable, and personality-consistent across diverse scenarios.

\section{Experiments}
We conduct experiments across three social contexts to evaluate ASVO's effectiveness in modeling human-like, socially-motivated multi-agent interactions. We assess behavioral naturalness, SVO adaptation dynamics, and emergent cooperation patterns against baseline approaches. Code and supplementary materials are available at \url{https://asvo-agents.github.io/ASVO-agents/}.

\subsection{Experimental Setups}
All experiments are conducted within an extended Concordia-based multi-agent simulation platform~\cite{wang2024concordia}, specifically designed to systematically evaluate the effects of dynamic SVO adaptation and desire-driven decision-making in diverse social contexts. The core simulation platform robustly supports flexible agent configuration, real-time interaction logging, and reproducible environment initialization for consistent and comparative experimental analysis.

\textbf{Environments and Scenarios.  }
Building on the established micro–meso–macro stratification paradigm~\cite{coleman1990foundations, epstein1996growing}, we extend the simulation framework beyond educational settings to encompass three representative \textbf{social contexts}, \textit{school}, \textit{family}, and \textit{workplace}. Each context contains multiple \textbf{social scales} (dyadic, small group, and large group), allowing the investigation of SVO adaptation and motivational reasoning under progressively complex social structures. The family context includes the first two scales, reflecting its naturally smaller social scope. Each context represents a distinct social type, where Family reflects small-scale interactions, School represents structured peer environments, and Workplace corresponds to institutionalized collaboration and competition.

   \textbf{\textit{School. }}  
    \textit{a) Micro:} A dormitory-like environment where two students share daily routines and emotions, focusing on interpersonal adjustment and empathy formation in close relationships.  
    \textit{b) Meso:} A classroom small-group setting in which students collaborate or compete for limited resources, reflecting the balance between cooperation, self-interest, and social comparison.  
    \textit{c) Macro:} A class election involving public speeches, alliances, and peer influence, capturing collective norm formation, leadership emergence, and competitive persuasion dynamics.
    
    \textbf{\textit{Workplace. }}  
    \textit{a) Micro:} A two-person office dyad reflecting hierarchical or peer communication under performance pressure, examining how SVO affects conflict management and support-seeking behaviors.  
    \textit{b) Meso:} A project team emphasizing collaboration, task division, and strategic coordination, where agents must negotiate responsibility and recognition.  
    \textit{c) Macro:} A departmental review simulating institutionalized cooperation and competition driven by evaluation, incentives, and career-oriented motivation, revealing how social values adapt to structured hierarchies.
    
    \textbf{\textit{Family. }}  
    \textit{a) Micro:} A weekend home scenario with two siblings staying indoors while parents are away, reflecting mild autonomy, cooperation, and self-regulation in intimate family interaction.  
    \textit{b) Meso:} A family park outing with siblings and parents engaging in shared play, small conflicts, and emotional bonding, highlighting the role of affection and social harmony in value adaptation.

This hierarchical and cross-context design enables a unified examination of how SVO-driven adaptation and desire-based reasoning evolve from intimate relationships to institutionalized cooperation and competition.
It also allows comparison of emergent behavioral patterns across social scales and domains, bridging micro-level interpersonal dynamics and macro-level collective outcomes.

\begin{figure}[t]
\vspace{-4pt}
\centering
\includegraphics[width=0.95\linewidth,trim=0 18 0 0,clip]{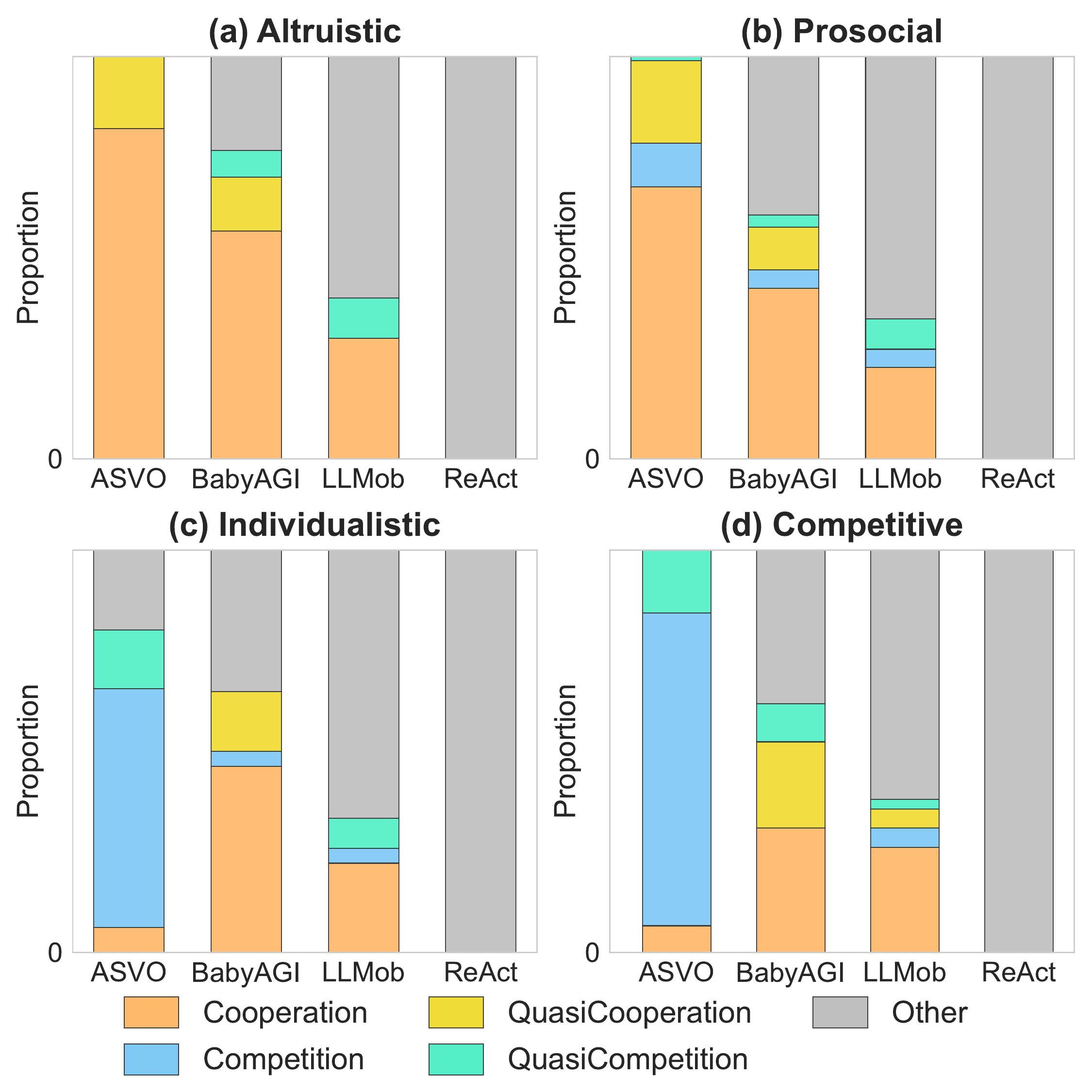}
\vspace{-4pt}
\caption{Behavioral distributions of four SVO-based social personality types, \textit{Altruistic}, \textit{Prosocial}, \textit{Individualistic}, and \textit{Competitive}, showing proportions of cooperation, competition, and intermediate behaviors across frameworks.}
\label{fig:svo-types}
\vspace{-12pt}
\end{figure}

\textbf{Language Model Selection.  }  
To comprehensively evaluate the impact of foundational models on multi-agent social simulation, we selected four representative large language models covering a range of scales and architectures: \textbf{GPT-5}~\cite{openai2024GPT4}, \textbf{Qwen3-235b}~\cite{yang2025Qwen3}, \textbf{Deepseek-v3}~\cite{Deepseek2024}, and \textbf{Gemini-2.5}~\cite{Geminiteam2025Gemini}. These models include both open-source and proprietary solutions and represent state-of-the-art advancements in LLMs. All models are utilized both as behavior generators (i.e., simulating agent actions and interactions) and as automatic evaluators (i.e., rating the naturalness and human-likeness of agent behaviors) within our experimental framework. For each experimental run, the same set of agent profiles, environmental contexts, and evaluation prompts is used to ensure a fair and reproducible comparison across LLMs.

\textbf{Evaluation Metrics.} We focus on the following key metrics to evaluate agent behavior and social adaptation:
1) \textbf{Cooperation and competition rate}: The frequency and proportion of cooperative and competitive actions among agents, as defined by whether agents' actions are promotive of or oppositional to others' goal attainment~\cite{johnson2005new}. 
2) \textbf{Naturalness}: The plausibility and human-likeness of agent behaviors, automatically evaluated by large language models using standardized prompts and scoring rubrics~\cite{welleck-etal-2019-dialogue}.
3) \textbf{Human-likeness}: The perceived similarity of agent behaviors to real human actions, as judged by LLM evaluators based on their conformity to human norms, diversity, and context-appropriateness~\cite{10.1145/3586183.3606763}.

\subsection{Quantitative Results}

To evaluate effectiveness, we compare ASVO with representative LLM-based agents, including ReAct~\cite{yao2023react}, BabyAGI~\cite{nakajima2023babyagi}, LLMob~\cite{wang2024a}, and D2A~\cite{wang2025desire} (an earlier intrinsic-motivation-driven framework), under the same simulation settings with matched profiles, goals, and context.
Only ASVO jointly models SVO and dynamic desires; the baselines follow their original designs (e.g., goal-driven planning, fixed habits, or static traits) without explicit SVO modeling.

Figure~\ref{fig:svo-types} illustrates the consistent and interpretable behavioral patterns of ASVO across the full spectrum.
Specifically, agents initialized with altruistic or prosocial profiles demonstrate a distinct tendency towards cooperative behaviors.
This validates the model's effective alignment with socially oriented motivations.
Conversely, as the SVO angle shifts toward individualistic orientations, the behavioral distribution transitions accordingly.
Cooperative tendencies diminish, while competitive actions become increasingly dominant.
This smooth, monotonic progression underscores ASVO’s capability to model the continuous behavioral plasticity inherent in human social values.

\begin{table}[t]
\small
\footnotesize
\setlength{\tabcolsep}{2pt}
\centering
\vspace{-6pt}
\caption{
Average \textbf{naturalness} (N) and \textbf{human-likeness} (H) scores for each LLM and method. 
Avg: mean value; Std: standard deviation across all LLMs for each method.
}
\vspace{-6pt}
\begin{tabular}{l|cc|cc|cc|cc|cc}
\hline
\multirow{2}{*}{LLM}
  & \multicolumn{2}{c}{ReAct}
  & \multicolumn{2}{c}{BabyAGI}
  & \multicolumn{2}{c}{LLMob}
  & \multicolumn{2}{c}{D2A}
  & \multicolumn{2}{c}{ASVO} \\
  & N & H & N & H & N & H & N & H & N &H \\
\cline{3-10}
\hline
\multicolumn{10}{l}{\textbf{School}} \\
\hline
Deepseek   & 4.004 & 3.667 & 3.438 & 2.771 & 4.500 & 3.708 & 3.292 & 3.083 & \textbf{4.750} & \textbf{4.792} \\
GPT-5      & 3.750 & 3.417 & 3.833 & 3.521 & 4.354 & 4.271 & 3.812 & 3.448 & \textbf{4.958} & \textbf{4.958} \\
Gemini-2.5     & 3.833 & 3.896 & 3.479 & 3.042 & 4.312 & 4.062 & 3.427 & 3.083 & \textbf{4.708} & \textbf{4.708} \\
Qwen3      & 3.917 & 3.521 & 3.458 & 3.188 & 4.271 & 4.021 & 3.677 & 3.406 & \textbf{4.792} & \textbf{4.824} \\
\hline
Avg        & 3.876 & 3.625 & 3.552 & 3.130 & 4.359 & 3.938 & 3.552 & 3.255 & \textbf{4.802} & \textbf{4.821} \\
Std        & 1.097 & 0.658 & 1.322 & 0.877 & 0.861 & 0.565 & 1.426 & 1.047 & \textbf{0.497} & \textbf{0.456} \\
\hline
\multicolumn{10}{l}{\textbf{Workplace}} \\
\hline
Deepseek   & 3.214 & 2.385 & 2.455 & 2.397 & 3.253 & 2.609 & 2.849 & 2.479 & \textbf{4.766} & \textbf{4.119} \\
GPT-5      & 3.221 & 3.045 & 3.349 & 3.064 & 4.144 & 3.792 & 3.703 & 3.458 & \textbf{4.849} & \textbf{4.019} \\
Gemini-2.5      & 3.448 & 3.182 & 2.673 & 2.724 & 3.442 & 3.269 & 2.891 & 2.698 & \textbf{4.878} & \textbf{4.034} \\
Qwen3      & 3.297 & 2.760 & 2.843 & 2.673 & 3.330 & 3.058 & 3.135 & 2.818 & \textbf{4.782} & \textbf{4.026} \\
\hline
Avg        & 3.330 & 2.877 & 2.798 & 2.710 & 3.542 & 3.182 & 3.145 & 2.863 & \textbf{4.819} & \textbf{4.049} \\
Std        & 1.568 & 1.063 & 1.358 & 0.968 & 1.463 & 1.088 & 1.444 & 0.991 & \textbf{0.491} & \textbf{0.466} \\
\hline
\multicolumn{10}{l}{\textbf{Family}} \\
\hline
Deepseek   & 3.286 & 3.262 & 2.917 & 2.690 & 3.589 & 3.339 & 4.010 & 3.167 & \textbf{4.750} & \textbf{3.964} \\
GPT-5      & 3.345 & 3.202 & 3.345 & 3.202 & 4.202 & 3.863 & 4.479 & 3.750 & \textbf{4.905} & \textbf{4.024} \\
Gemini-2.5 & 3.726 & 3.488 & 3.423 & 1.433 & 3.845 & 3.429 & 4.031 & 3.615 & \textbf{4.863} & \textbf{4.077} \\
Qwen3      & 4.006 & 3.536 & 3.685 & 3.315 & 4.000 & 3.738 & 4.490 & 3.781 & \textbf{4.381} & \textbf{3.820} \\
\hline
Avg        & 3.693 & 3.449 & 3.342 & 3.079 & 3.909 & 3.579 & 4.253 & 3.578 & \textbf{4.725} & \textbf{3.946} \\
Std        & 1.423 & 0.921 & 1.239 & 0.904 & 1.166 & 0.827 & 1.148 & 0.941 & \textbf{0.624} & \textbf{0.648} \\
\hline
\end{tabular}
\label{tab:llm-method-row-col}
\vspace{-14pt}
\end{table}

In contrast, baseline models fail to reproduce this structured trend: their generated actions remain largely insensitive to SVO type, often demonstrating neutral or context-independent behavior.
ASVO uniquely demonstrates a well-ordered continuum of cooperation–competition balance and complex behavioral diversity, bridging quantitative SVO parameters with observable outcomes.

\begin{figure}[t]
\centering
\vspace{-4pt}
\includegraphics[width=0.98\linewidth,trim=0 21 0 25,clip]{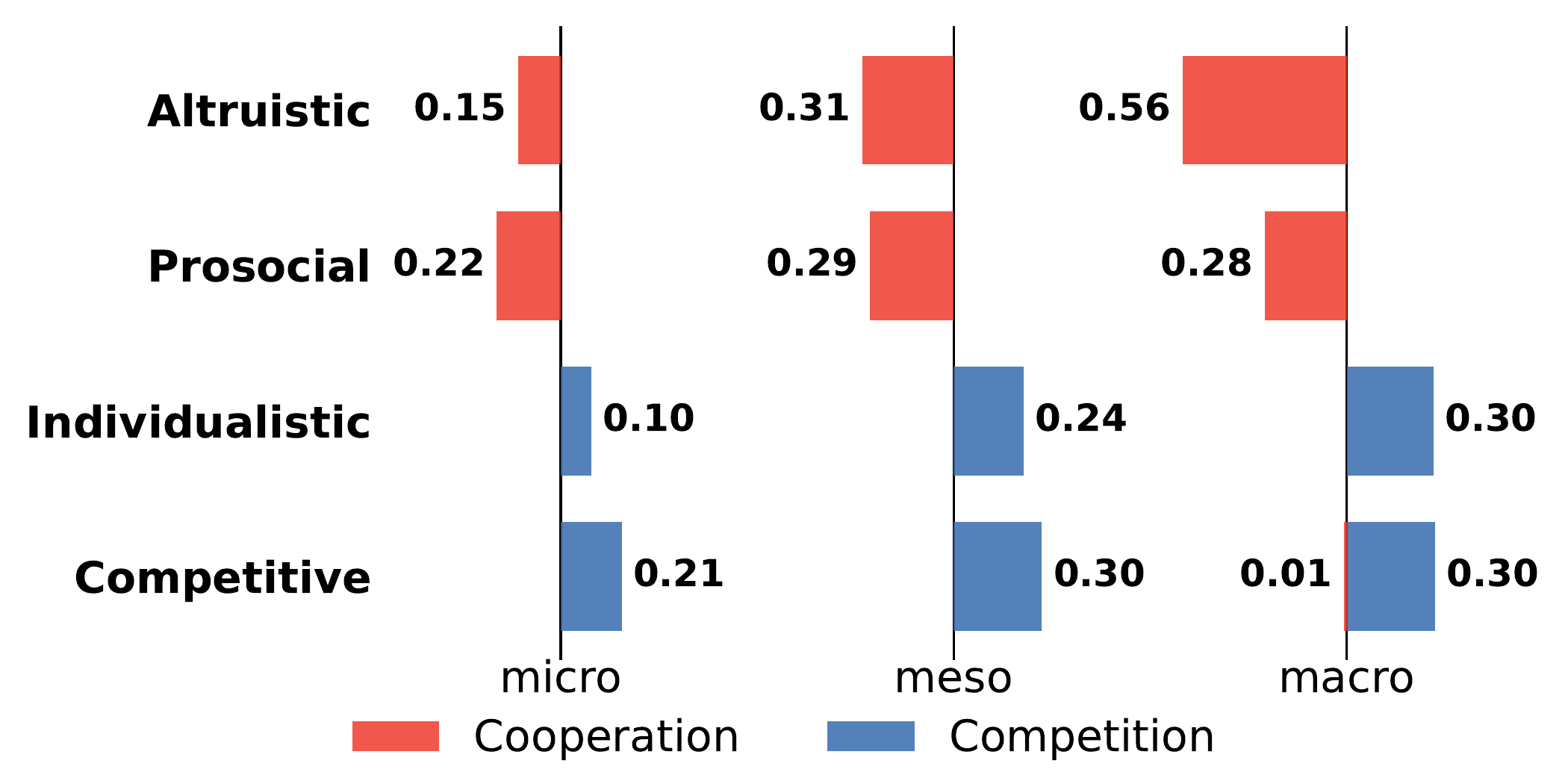}
\vspace{-4pt}
\caption{Distribution of cooperative and competitive actions for each social personality type within the \textit{School} context across micro-, meso-, and macro-level scales.}
\vspace{-15pt}
\label{fig:mirror-bar}
\end{figure}

Beyond capturing this graded behavioral spectrum, we evaluate the \textit{naturalness} and \textit{human-likeness} of generated behaviors across three social contexts: \textit{School}, \textit{Workplace}, and \textit{Family}. As shown in Table~\ref{tab:llm-method-row-col}, \textbf{ASVO} consistently achieves the highest scores across all contexts and LLM backbones, with remarkably low standard deviations indicating strong robustness and generalization. The smaller performance gaps among different LLMs suggest that value-driven adaptation effectively mitigates backbone variability. Multi-judge stability analysis is provided in Appendix~D.4. 
These results confirm that incorporating explicit value structures and adaptive SVO mechanisms substantially enhances behavioral coherence and social consistency in multi-agent simulations, enabling agents to express distinct yet contextually appropriate social behaviors.

\begin{figure}[t]
\centering
\includegraphics[width=0.9\linewidth,trim=0 17 0 24,clip]{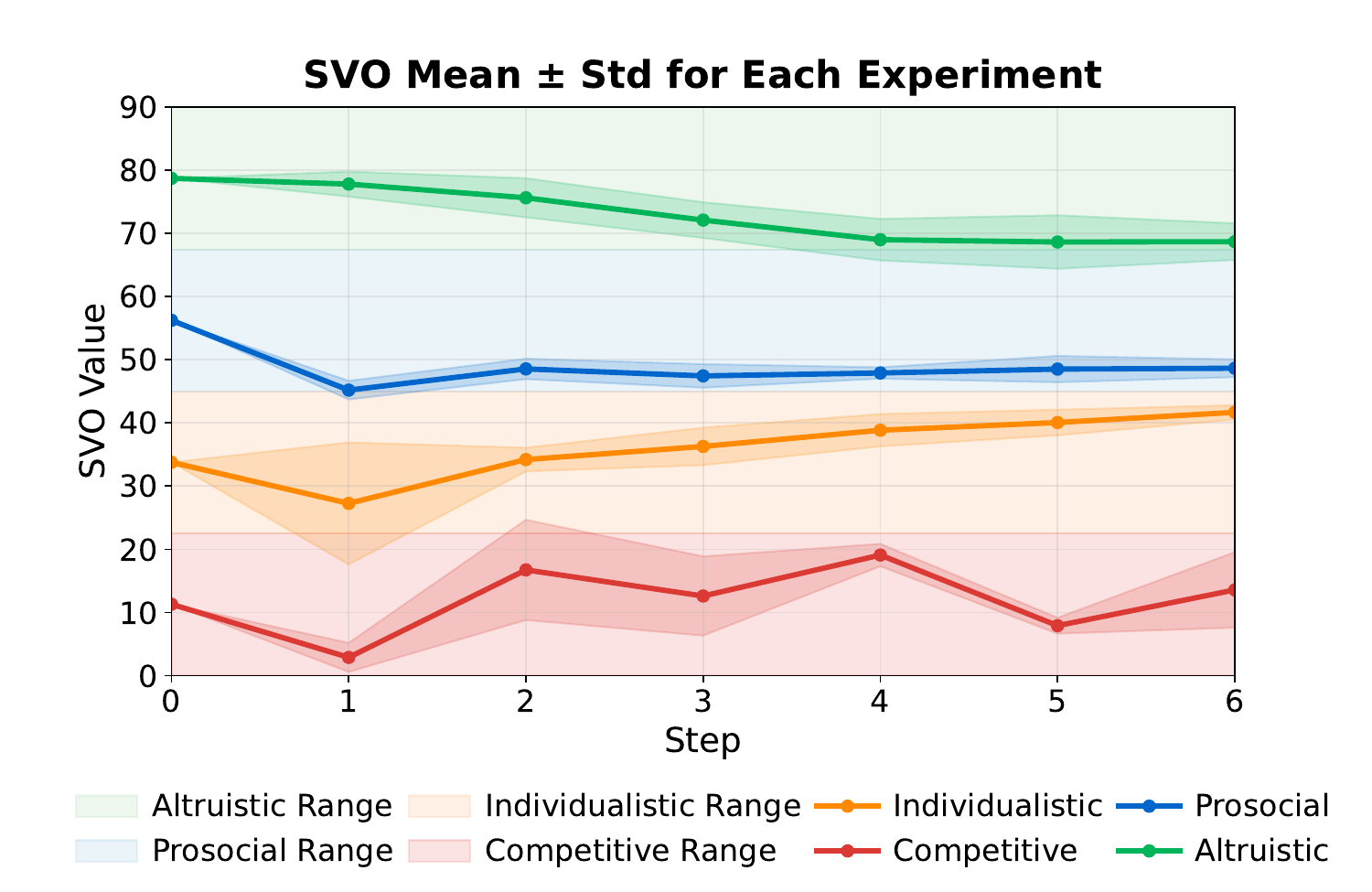}
\vspace{-4pt}
\caption{Temporal evolution of SVO mean and standard deviation for each personality type. Shaded areas denote the reference range of each SVO category, with values remaining within their respective intervals across simulation steps.}
\vspace{-4pt}
\label{fig:svo-range}
\end{figure}

To further assess the fidelity of our SVO modeling, we further examine whether agents' SVO values remain within their intended theoretical ranges over time. As illustrated in Figure~\ref{fig:svo-range}, the mean and standard deviation of SVO for each personality type are plotted alongside shaded bands representing the expected SVO intervals. Across all steps, altruistic, prosocial, individualistic, and competitive agents consistently maintain SVO values within their respective reference ranges. This result confirms that ASVO effectively preserves distinct social value orientations throughout the simulation.

\subsection{Scalability Analysis}
To test whether ASVO can scale to larger scenarios, we extend the experiments by varying the horizon length and the agent population size. We use the \textit{School} setting as a representative testbed to keep the environment fixed while changing only the scale factors (Table~\ref{tab:scale-robustness}).
We extend the simulation horizon from $6$ to $12$, $18$, and $24$ steps. ASVO maintains coherent and human-like behaviors as the horizon grows, with degradation at $24$ steps.
We increase the population from $4$ to $8$, $16$, and $32$ agents under the same setting. Scores decrease gradually with more agents, indicating ASVO remains effective when scaling to larger groups.
\begin{table}[t]
\small
\footnotesize
\setlength{\tabcolsep}{2pt}
\centering
\vspace{-2pt}
\caption{
Robustness of ASVO under extended temporal horizons and agent population sizes in the \textit{School} setting.
}
\vspace{-4pt}
\begin{tabular}{c|cc|cc|cc|cc|cc}
\hline
\multirow{2}{*}{Setting}
  & \multicolumn{2}{c}{Altruistic}
  & \multicolumn{2}{c}{Prosocial}
  & \multicolumn{2}{c}{Individualistic}
  & \multicolumn{2}{c}{Competitive}
  & \multicolumn{2}{c}{Avg} \\
  & N & H & N & H & N & H & N & H & N & H \\
\cline{3-10}
\hline
\multicolumn{11}{l}{\textbf{Extended Temporal Horizons (Steps)}} \\
\hline
6   & 4.83 & 4.33 & 4.83 & 4.33 & 4.50 & 4.20 & 4.33 & 4.30 & 4.63 & 4.29 \\
12  & 5.00 & 4.51 & 5.00 & 4.17 & 4.75 & 4.50 & 4.25 & 4.42 & 4.75 & 4.39 \\
18  & 4.89 & 4.39 & 4.94 & 4.50 & 4.67 & 4.23 & 4.78 & 3.83 & 4.82 & 4.23 \\
24  & 4.42 & 4.21 & 4.83 & 4.50 & 4.46 & 4.29 & 4.29 & 4.08 & 4.50 & 4.27 \\
\hline
Avg & 4.79 & 4.36 & 4.90 & 4.38 & 4.60 & 4.31 & 4.41 & 4.16 & 4.68 & 4.30 \\
\hline
\multicolumn{11}{l}{\textbf{Increased Agent Population Sizes (Agents)}} \\
\hline
4   & 4.83 & 4.33 & 4.83 & 4.33 & 4.50 & 4.23 & 4.33 & 4.30 & 4.63 & 4.29 \\
8   & 4.75 & 4.43 & 4.71 & 4.33 & 4.20 & 4.25 & 4.17 & 4.33 & 4.56 & 4.37 \\
16  & 4.79 & 4.43 & 4.80 & 4.45 & 4.16 & 4.23 & 4.58 & 4.31 & 4.58 & 4.33 \\
32  & 4.60 & 4.15 & 4.35 & 4.21 & 3.92 & 4.25 & 4.35 & 4.21 & 4.42 & 4.27 \\
\hline
Avg & 4.74 & 4.34 & 4.67 & 4.33 & 4.20 & 4.24 & 4.36 & 4.29 & 4.55 & 4.32 \\
\hline
\end{tabular}
\label{tab:scale-robustness}
\vspace{-6pt}
\end{table}

Beyond quantitative metrics, longer-horizon simulations demonstrate clearer long-range behavioral continuity.
Altruistic agents gradually evolve from basic peer assistance to more structured cooperative behaviors, such as monitoring peers’ progress, organizing study groups, and offering emotional support.
Competitive agents increasingly focus on performance disparities and may show frustration when others progress faster, while individualistic agents consistently favor independent and self-directed learning.
In contrast, prosocial agents progressively converge toward stable supportive roles as group dynamics unfold.

\subsection{Behavioral Patterns Across Environments}
To investigate how social personality shapes agent behavior across social contexts, we conducted controlled experiments in which all agents within each environment were assigned the same personality type. Each scenario was repeated five times with six simulation steps per run, and agent actions were aggregated for analysis.

Figure~\ref{fig:mirror-bar} shows how cooperation and competition rates vary with personality type and social scale within the \textit{School}.
Altruistic and prosocial agents consistently demonstrate higher proportions of cooperative actions, whereas individualistic and competitive agents are more prone to competitive behaviors.
As the environment expands from micro (dormitory) to meso (classroom) and macro (class election), cooperation rates increase across all personality types, for example, from $0.15$ to $0.56$ for altruistic agents and from $0.22$ to $0.28$ for prosocial agents.
In contrast, competition rates show subtler changes: they rise slightly for prosocial and competitive agents but decline marginally for altruistic and individualistic ones.
This pattern indicates that larger and more complex environments enhance overall social engagement, leading agents to demonstrate richer and more differentiated behavioral patterns rather than uniformly greater competitiveness.


In particular, altruistic agents show the strongest context sensitivity, with cooperation rates more than tripling from the dormitory to the class election scenario. This pattern suggests that altruistic SVOs are highly responsive to expanded social visibility and collective incentives, leading to cooperative behavior in larger peer groups.
Prosocial agents maintain a stable balance between self- and other-oriented motivations, demonstrating modest and consistent cooperation growth as environments scale up.
By contrast, individualistic and competitive agents show lower cooperation ratios, with increased behavioral variance in meso- and macro-level contexts, reflecting adaptive shifts between self-interest and strategic social positioning.
Overall, these findings show that ASVO captures flexible value-driven decision-making, with cooperative and competitive tendencies shaped by personality and social structure.

\begin{figure}[t]
    \centering
    \includegraphics[width=0.82\linewidth,trim=0 28 0 10,clip]{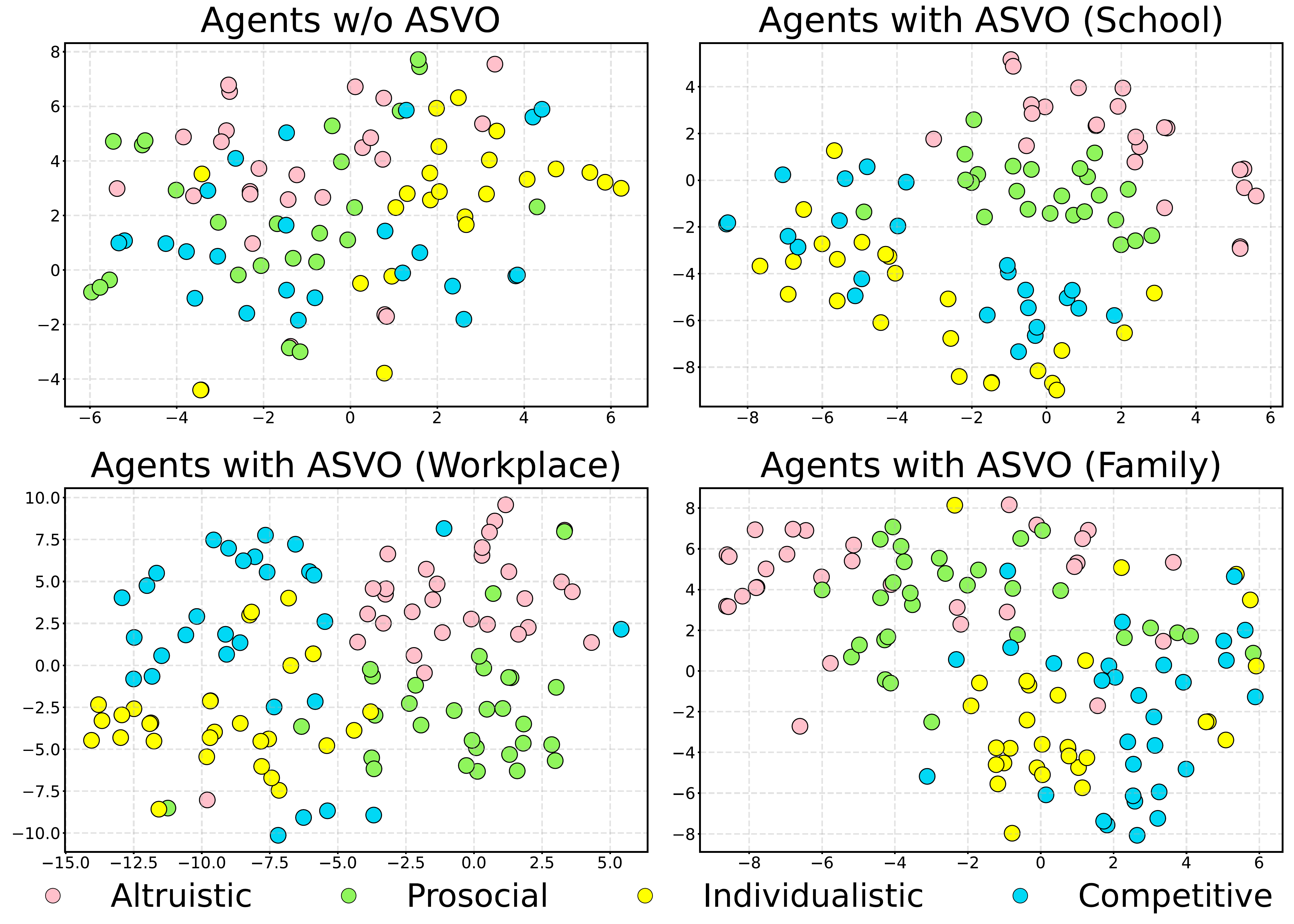}
    \vspace{-4pt}
    \caption{t-SNE visualization of agent behavior embeddings across four SVO personality types. The top-left panel shows agents without SVO, while the others demonstrate our agents (ASVO) across \textit{School}, \textit{Workplace}, and \textit{Family}. ASVO yields clearer clustering, indicating more disentangled behaviors.
}
    \label{fig:tsne-ablation}
    \vspace{-12pt}
\end{figure}

\subsection{Ablation Studies of the LLMs Choice}
To systematically assess the impact of different LLMs on emergent social behaviors across multiple social contexts, we conducted experiments under three representative environments: \textit{School}, \textit{Workplace}, and \textit{Family}. We employed both \textbf{naturalness} and \textbf{human-likeness} as automatic evaluation metrics to measure the fluency and human-level coherence of agent behaviors.

Table~\ref{tab:llm-method-row-col} presents the average scores obtained by each LLM across the three contexts. The results reveal clear performance distinctions among models. \textbf{GPT-5} and \textbf{Qwen3} achieve the highest overall scores in both metrics, indicating their stronger capability in generating coherent, socially aligned, and context-sensitive behaviors. In contrast, \textbf{Deepseek} and \textbf{Gemini} show greater variability across contexts, reflecting higher sensitivity to environmental complexity and interaction structure.
Among the three environments, the \textit{Workplace} context shows the highest variability, reflecting the increased complexity and role diversity of organizational interactions. In contrast, the \textit{Family} context yields lower overall scores due to its emotionally driven and less structured interactions, while the \textit{School} context maintains balanced performance across all models. These results collectively highlight that both the social setting and the underlying LLM jointly shape the emergence of realistic and personality-consistent behaviors in large-scale simulations.

To evaluate whether ASVO enables the emergence of consistent personality-driven behaviors, we performed an ablation study using t-SNE to visualize the action distribution of four personality types. By comparing the baseline without ASVO and the ASVO-enabled agents across three social contexts (School, Workplace, and Family), we directly observe how explicit value orientation modeling enhances the separability of agent behaviors.

In the absence of ASVO (top left), the actions of all personality types are highly intermixed in the t-SNE space, indicating that baseline agents lack consistent alignment between personality and behavior. There is no clear cluster structure, suggesting that static or shallow modeling fails to capture or preserve individual value orientations during social interactions. Figure~\ref{fig:tsne-ablation} shows that, with ASVO, the actions of Altruistic and Prosocial agents are clearly separated from those of Individualistic and Competitive agents across all three social contexts (School, Workplace, and Family), reflecting robust personality-driven clustering.

This result clearly demonstrates that integrating structured social value orientation not only increases the interpretability and realism of multi-agent behavior, but also substantially enhances the separation and robustness of personality-driven dynamics.
Because t-SNE is a qualitative visualization and may appear less distinctive when embeddings overlap, we additionally quantify cluster separability using the silhouette score and a cluster separation ratio computed from the same behavior embeddings. Detailed results are provided in the Appendix~C.

\section{Conclusion and Future Work}
This paper presents the \textbf{Autonomous Social Value-Oriented agents (ASVO)} framework, which integrates desire-driven autonomy with dynamically evolving Social Value Orientation (SVO).
By combining intrinsic motivational reasoning with adaptive value regulation, ASVO enables LLM agents to demonstrate socially adaptive and personality-consistent behaviors across diverse contexts.
Experiments across school, workplace, and family environments show that ASVO reproduces key social phenomena, including cooperation, competition, and value alignment, consistent with SVO theory.
The framework offers a structured foundation for studying value-driven decision-making and supports applications in education, social training, and human-AI interaction.

\textbf{Limitations: }Despite its promising results, ASVO has two main limitations.
First, the current motivational system relies on a fixed set of core desires.
This design cannot fully capture the richness and variability of human values.
Second, the simulated environments remain relatively controlled and deterministic.
Although structurally diverse, they limit the observation of emergent adaptation in open-ended or uncertain social contexts.

\textbf{Future Work: }We will extend the internal value system to support self-evolving motivational structures during simulation.
We also plan to explore more dynamic, open, and cross-cultural environments in which agents infer roles, negotiate norms, and adapt to complex social dynamics.
In addition, incorporating multimodal signals may further improve the realism and complexity of agent interactions.

\section*{Acknowledgments}
This work was supported by Beijing Natural Science Foundation L252010, NSFC-62406010, and the Fundamental Research Funds for the Central Universities.


\bibliographystyle{ACM-Reference-Format} 
\bibliography{sample}


\end{document}